\documentclass[epj,nopacs,a4paper]{mysvjour}

\usepackage{epsfig}
\usepackage{amsmath}
\usepackage{array}
\usepackage{axodraw}
\usepackage{cite}

\newcommand{\qbar}{\overline{q}}
\newcommand{\pom}{I\!\!P}
\newcommand{\xpom}{x_{I\!\!P}}
\newcommand{\zpom}{z_{I\!\!P}}
\newcommand{\reg}{I\!\!R}
\newcommand{\PreserveBackslash}[1]{\let\temp=\\#1\let\\=\temp}

\begin{document}

 \title{Final States in Diffraction at HERA\thanks{Talk given at the Crimean Summer School-Seminar {\it New Trends in High-Energy Physics}, May 27--June 4 2000, Yalta (Miskhor), Ukraine}}

\author{Pierre Van Mechelen\thanks{Postdoctoral fellow of the Fund for Scientific Research - Flanders (Belgium)}}

\institute{University of Antwerpen (UIA), Belgium; \email{Pierre.VanMechelen@uia.ua.ac.be}}

\abstract{Recent experimental results on diffractive high $E_T$ jet production and diffractive open charm production, obtained by the H1 and ZEUS experiments at HERA, are presented.  Various theoretical models for diffractive hard scattering are discussed and confronted with the experimental data.}

\date{ }

\maketitle

\section{Introduction}
\label{sec:introduction}

It is well known that soft hadron-hadron scattering can be described by Regge phenomenology, which models the total, elastic and single and double diffractive dissociation cross sections in terms of reggeon and pomeron exchange \cite{bib:collins_regge_theory}.  At high energies, it is the pomeron exchange mechanism that dominates the total cross section \cite{bib:dl}.  However, until recently, our understanding of the pomeron in terms of Quantum Chromodynamics (QCD) \cite{bib:forshaw_qcdpomeron} remained fragmentary, at best.

The observation of events with a large rapidity gap in the hadronic final state at HERA \cite{bib:hera_lrg_obs}, which are attributed to diffractive dissociation of (virtual) photons, led to a renewed interest in the study of the underlying dynamics of diffraction.  In deep inelastic $ep$ scattering, the long hadronic lifetime of the photon at small Bjorken-$x$ and its variable virtuality $Q^2$ allow to make the link with diffractive dissociation in hadron-hadron scattering, while the presence of one or more hard scales enables perturbative calculations in QCD.

In the framework of QCD, it is tempting to attribute a partonic structure to the pomeron \cite{bib:ingelman_schlein}; an approach which proves to be very successful in describing various aspects of diffractive virtual photon dissociation. QCD calculations indicate, however, that the pomeron does not exhibit a universal behaviour over the full kinematic range. Indeed, experimental observations of a transition do exist when going from the soft to the hard scattering regime, changing the effective intercept of the pomeron trajectory from 1.08 in the soft limit to 1.2 in harder interactions. 

Early experimental measurements at HERA of the inclusive cross section \cite{bib:hera_incl_ddis,bib:h1_reggeqcd_fits,bib:zeus_diffdis_ratio} led to the conclusion that a partonic pomeron would be dominated by hard gluons.  This has been qualitatively confirmed by several studies of the properties of the hadronic final state of diffractive virtual photon dissociation \cite{bib:hera_diff_hfs}, which are all in agreement with the fragmentation of a colour octet-octet partonic final state, leading to higher particle multiplicities, lower thrust values and a a final state system that is not strictly aligned with the $\gamma^\ast$ axis in the rest frame of the photon dissociation system.

This article presents some recent experimental measurements of more exclusive final states\footnote{Previous results on jet production in diffractive deep inelastic scattering can be found in \cite{bib:hera_diff_excl_hfs}.}. The production of open charm or of jets with large transverse energy $E_T$ is particularly sensitive to the gluon content of the pomeron. Moreover, the high $E_T$ and large charm quark mass provide an additional hard scale for perturbative QCD calculations. Following this introduction, Sec.\@~\ref{sec:kinematics} provides a short reminder of the kinematics of diffractive deep inelastic scattering.  Section \ref{sec:theory} discusses the theoretical models that are used for comparison with the experimental results, Sec.\@~\ref{sec:experiment} presents the preliminary results, obtained by the H1 and ZEUS collaborations, on high $E_T$ dijet production and open charm production and, to conclude, a summary is given in Sec.\@~\ref{sec:summary}.

\section{Kinematics of diffractive \boldmath{$ep$} scattering}
\label{sec:kinematics}

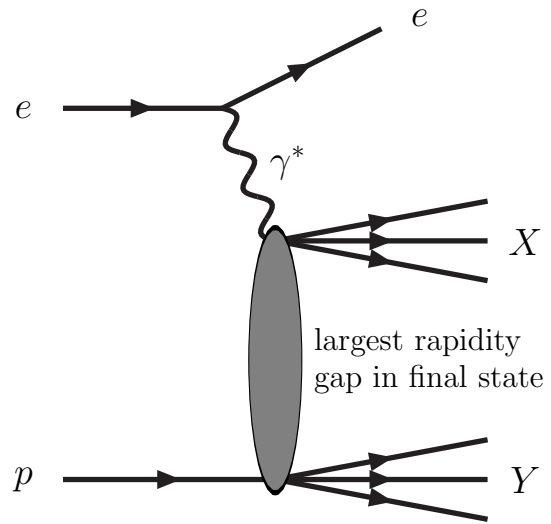
\begin{figure}
\begin{large}
\begin{center}
\begin{picture}(200,210)
\SetWidth{2}
\ArrowLine(20,35)(100,35)       \Text(5,35)[c]{$p$}
\ArrowLine(100,35)(180,50)
\ArrowLine(100,35)(180,35)      \Text(195,35)[c]{$Y$}
\ArrowLine(100,35)(180,20)
\ArrowLine(20,175)(80,175)      \Text(5,175)[c]{$e$}
\ArrowLine(80,175)(140,205)     \Text(155,210)[c]{$e$}
\Photon(80,175)(100,125){4}{3}  \Text(105,155)[c]{$\gamma^\ast$}
\ArrowLine(100,125)(180,140)
\ArrowLine(100,125)(180,125)    \Text(195,125)[c]{$X$}
\ArrowLine(100,125)(180,110)
\GOval(100,80)(50,10)(0){0.5}   \Text(115,87)[l]{\normalsize largest rapidity}
                                \Text(115,73)[l]{\normalsize gap in final state}
\end{picture}
\end{center}
\end{large}
\caption{Illustration of the generic diffractive process $ep \rightarrow eXY$.}
\label{fig:kinematics}
\end{figure}

The generic diffractive process $ep \rightarrow eXY$ is illustrated in Fig.\@~\ref{fig:kinematics}. A photon with virtuality $Q^2=-q^2$ interacts with the proton to produce the two final state systems $X$ and $Y$, which are separated by a large rapidity gap if their invariant masses are small.  Cross sections can be expressed in terms of $Q^2$ and the invariant masses of the $\gamma^\ast p$ system ($W$), and the $X$ and $Y$ systems ($M_X$ and $M_Y$, resp.\@). In addition, the following variables are usually introduced:
\begin{align}
&\xpom = \frac{q \cdot (p - p_Y)}{q \cdot p} \approx \frac{Q^2 + M_X^2}{Q^2 + W^2}, \\
&\beta = \frac{Q^2}{2 q \cdot (p - p_Y)} \approx \frac{Q^2}{Q^2 + M_X^2}.
\end{align}
In the proton infinite momentum frame, $\xpom$ is the fraction of the beam proton momentum transferred to $X$, while $\beta$ represents the fraction of the exchanged momentum carried by the struck quark.  The squared four-momentum transferred at the proton vertex ($t$) has been neglected in the above equations.

\section{Theoretical models}
\label{sec:theory}

Theoretical models for diffractive deep inelastic scattering range from purely phenomenological parameterizations with a large number of degrees of freedom to full perturbative QCD calculations. The following briefly discusses those models that are used for comparisons to experimental data.

\subsection*{The partonic pomeron} 

When measuring the diffractive cross section in deep inelastic scattering at HERA, it was soon recognized that the $\xpom$ dependence can be modeled exploiting Regge factorization \cite{bib:gribov_pomeranchuk}, provided that a sub-leading exchange is allowed to contribute at high values of $\xpom$.  In this model, the diffractive cross section factorizes into pomeron and reggeon flux factors, depending on $\xpom$ only, and into pomeron and reggeon structure functions, depending on $\beta$ and $Q^2$. The flux factors describe the long distance physics at the proton vertex and can be parameterized using Regge theory, while the structure functions are free parameters containing the short distance physics at the photon vertex.  

Recently, QCD factorization has been proven to hold for the leading twist component of diffractive deep inelastic scattering \cite{bib:collins_diff_qcdfact}, allowing to further interpret the structure functions in terms of quarks and gluons by applying a QCD motivated model for their $\beta$ and $Q^2$ dependence \cite{bib:ingelman_schlein,bib:diff_pdf}. Parton density functions for the pomeron have thus been extracted from the measured cross section, assuming the pion structure function for the sub-leading reggeon exchange and the next-to-leading order DGLAP equations \cite{bib:dglap} for the evolution in $Q^2$.

The outcome of these combined Regge-QCD fits is a value of around 1.2 for the pomeron intercept $\alpha_{\pom}(0)$ and a pomeron structure function that is dominated by gluons carrying large fractional momenta \cite{bib:h1_reggeqcd_fits}.

\subsection*{Soft colour interaction model}

A model that does not use the notion of a pome\-ron at all, is the soft colour interaction model \cite{bib:sci}. Instead, the standard matrix element and parton shower description of deep inelastic scattering is used, which, at low Bjorken-$x$, is dominated by boson-gluon fusion.  In addition, non-perturbative interactions affect the final colour connections between partons, while leaving the parton momenta unchanged. This may result in an interruption of the colour strings between partons, thus creating a large rapidity gap in the final state.  The probability for such a soft colour interaction has to be fixed by the experimental data.  

Recently, an improved version of this model has become available, which is based on a generalised area law and which favours soft colour interactions that make the colour strings between partons shorter \cite{bib:sci_gal}.

\subsection*{Saturation model}

In this and the subsequent models, the virtual photon is considered to fluctuate in partonic configurations, like e.g.\@ $q\qbar$ or $q\qbar g$. In the proton rest frame and at low Bjorken-$x$, this happens long before the actual interaction with the proton.  Both states, $q \qbar$ and $q \qbar g$, can be described by dipole wave-functions.  These wave-functions are responsible for the basic shape of the $\beta$ dependence of the diffractive structure function. Three contributions have been introduced \cite{bib:bartels_photonfluct}, two of which are related to the production of $q\qbar$ pairs by longitudinally or transversely polarised photons, while a third contribution takes the radiation of an additional gluon into account for the case of a transversely polarised photon:
\begin{align}
&F_{q\qbar}^L \sim \frac{Q_0^2}{Q^2} \ln \left(\frac{Q^2}{4 Q_0^2 \beta}\right) \beta^3 (1 - 2 \beta)^2, \\
&F_{q\qbar}^T \sim \beta (1-\beta), \\ 
&F_{q\qbar g}^T \sim \alpha_S \ln \left(\frac{Q^2}{Q_0^2}\right) (1-\beta)^3.
\end{align}
From these formulae, it is clear that $q\qbar g$ states are expected to populate the low $\beta$ (i.e.\@ high mass) region, while $q\qbar$ states dominate at intermediate and high values of $\beta$.  Note that the longitudinal $q\qbar$ contribution is of higher twist nature and that the longitudinal cross section is expected to dominate over the transverse cross section at high $\beta$. 

The subsequent models differ in the way they treat the actual dipole-proton interaction.

An attempt to describe the saturation of the inclusive cross section at low $Q^2$ and low $x$ leads to the following parameterization of the dipole-proton cross section \cite{bib:saturation_model}:
\begin{equation}
\hat{\sigma} (x, r^2) = \sigma_0 \left[1 - \exp \left(-\frac{r^2}{4 R_0^2(x)}\right)\right],
\end{equation}
with
\begin{equation}
R_0(x) = \frac{1}{\rm GeV} \left(\frac{x}{x_0}\right)^{\lambda/2}.
\end{equation}
This dipole cross section exhibits colour transparency at small dipole radii $r$ (i.e.\@ $\hat{\sigma} \sim r^2$ for $r < R_0$) and saturation at large $r$ (i.e.\@ $\hat{\sigma} \sim \sigma_0$ for $r > R_0$).  Saturation at low $x$ is explicitly built-in through the $x$-dependent radius $R_0$.

After a fit to the inclusive structure function, the diffractive structure function follows as a prediction.

One interesting property of this model is the prediction that the ratio of diffractive to inclusive cross sections is roughly constant as a function of $W$, as has been observed experimentally \cite{bib:zeus_diffdis_ratio}.

\subsection*{Semi-classical model}

In the semi-classical approach \cite{bib:semiclassical}, an approximation is used in which the hadronic target is much larger than the transverse separation of the dipoles.  Assuming further that the colour fields in distant parts of the target are uncorrelated,  the relevant colour field configurations can be averaged, yielding explicit formulae for both the diffractive and inclusive parton distributions at some low scale $Q_0^2$.  Diffraction occurs if the emerging partonic state is a colour singlet. A conventional DGLAP analysis is then used to calculate the diffractive and inclusive structure functions at larger values of $Q^2$.  The model has four free parameters in total, which are obtained from a fit to the experimental data.

It is interesting to note that also in this model the diffractive gluon density turns out to be much larger than the diffractive quark density.

\subsection*{Two-gluon exchange model}

This QCD calculation describes the diffractive $\gamma^\ast p$ interaction in terms of partonic fluctuations of the photon ($q\qbar$, $q\qbar g$, etc.) that scatter elastically off the proton through the exchange of two gluons in a net colour-singlet configuration \cite{bib:twogluon}.  An essential feature of the model is that the cross section is proportional to the square of gluon density inside the proton.  The calculation requires high transverse momentum of all outgoing partons, thus making it particularly suited for diffractive jet production.  In order to avoid the valence quark region in the proton and to exclude secondary reggeon exchanges the range of validity is restricted to $\xpom < 0.01$.

\section{Experimental results}
\label{sec:experiment}

In the following, Monte Carlo models are used to compare the measured hadron level cross sections to the predictions of the phenomenological models and QCD calculations presented in the previous section.  RAPGAP \cite{bib:rapgap} is used to obtain the predictions of the partonic pomeron model with different pomeron intercept values and parton distributions.  The programme also contains implementations of the saturation and the semi-classical models as well as the two-gluon exchange model.  Both versions of the soft colour interaction model are implemented in the LEPTO generator \cite{bib:lepto}.

\begin{figure}[t]
\begin{center}
\epsfig{file=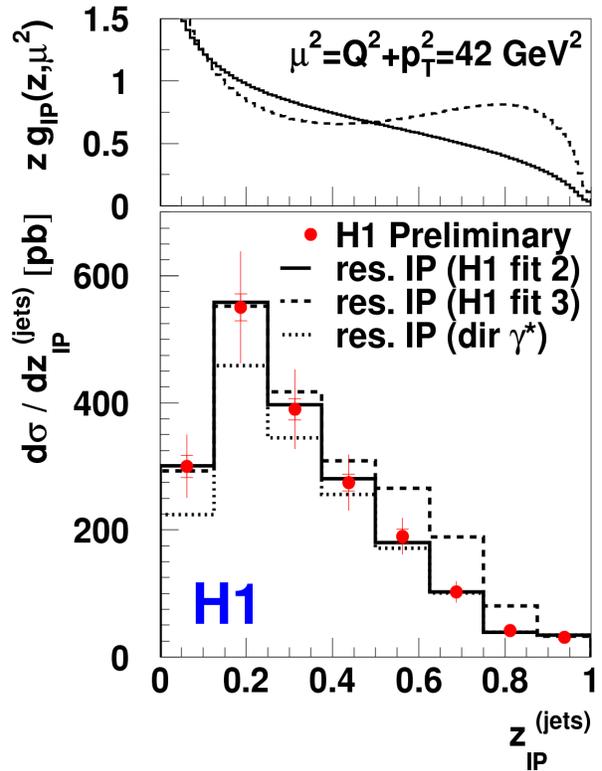,width=0.45\textwidth}
\end{center}
\caption{The differential diffractive dijet cross section is displayed as a function of $\zpom^{jets}$ and compared with predictions from the partonic (``resolved'') pomeron model.  Different gluon density functions, obtained from fits to the inclusive cross section and shown in the top figure for an evolution scale which is typical for the dijet event sample, are used as input to the model. The effect of only including the direct photon contribution is also shown.}
\label{fig:jets_zpom_partpom}
\end{figure}

\begin{table}
\begin{center}
\setlength{\extrarowheight}{2pt}
\begin{tabular}{c}
$4 < Q^2 < 80 {\rm\ GeV}^2$ \\
$0.1 < y < 0.7$ \\
\hline
$\xpom < 0.05$ \\
$M_Y < 1.6 {\rm\ GeV}$ \\
$|t| < 1 {\rm\ GeV}^2$ \\
\hline
$p_T^{jet} > 4 {\rm\ GeV}$ \\
$-3 < \eta^{jet} < 0$ \\
\end{tabular}
\end{center}
\caption{Kinematic domain of the diffractive dijet cross section measurement by H1.}
\label{tab:jetkine}
\end{table}

\subsection*{Diffractive high \boldmath{$E_T$} jet production}

\begin{figure*}
\begin{center}
\epsfig{file=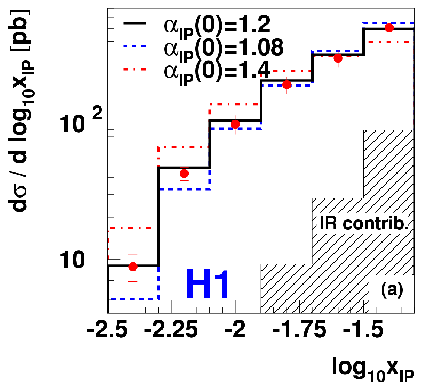,width=0.45\textwidth}
\hspace{0.05\textwidth}
\epsfig{file=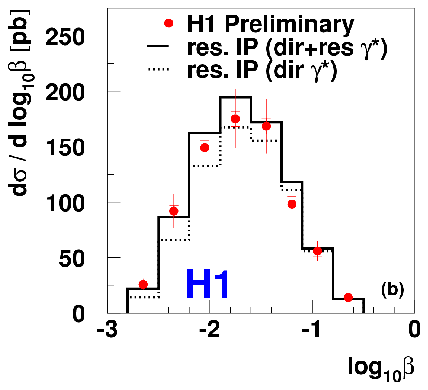,width=0.45\textwidth}
\end{center}
\caption{The differential dijet cross section is displayed as a function of $\xpom$ (a) and $\beta$ (b) and compared to the partonic pomeron model.  Predictions are shown for different values of the pomeron intercept $\alpha_{\pom}(0)$ and also the sub-leading reggeon component is displayed.}
\label{fig:jets_xpom_partpom}
\end{figure*}

The production of jets with large transverse energy provides an ideal test of the underlying dynamics of diffraction models \cite{bib:mcdermott_dijets}.  The interaction mechanism, which is dominated by boson-gluon fusion, is highly sensitive to the role of gluons and the large transverse energy lends a hard scale for perturbative calculations.

The H1 Collaboration recently obtained preliminary results based on an integrated luminosity of $17.9 {\rm\ pb}^{-1}$, resulting in approximately 2500 events with at least two jets in the final state system $X$.  Diffractive events are selected by excluding any hadronic activity in the range $3.2 < \eta^{lab} < 7.5$ and
jets are identified as collimated energy depositions in $\eta$-$\phi$ space within a cone of radius $R=\sqrt{\Delta \eta^2 + \Delta \phi^2}=1$.  The fraction of the pomeron momentum transferred to the dijet system, and therefore entering the hard scattering, can be calculated as follows:
\begin{equation}
z^{jets}_{\pom} = \frac{Q^2 + M_{jj}^2}{Q^2 + M_X^2}
\end{equation}
with $M_{jj}$ the invariant mass of the dijet system. Differential cross sections are measured in the kinematic region defined in Tab.\@~\ref{tab:jetkine} as a function of many variables \cite{bib:h1_future_dijets}.  Here, only a selection of the available distributions is presented.

Figure \ref{fig:jets_zpom_partpom} shows that applying the results of  Regge-QCD fits, as obtained from the diffractive structure function \cite{bib:h1_reggeqcd_fits}, works very well. The analysis of the inclusive diffractive cross section yields two different fits with similar $\chi^2/{\rm ndf}$ (cf.\@ inset on top of fig.\@~\ref{fig:jets_zpom_partpom}), each with very similar quark densities but one with a flat gluon density function (``H1 fit 2'') and another with a gluon density function that peaks at high values of the fractional momentum $z$ (``H1 fit 3'').  For the dijet cross section analysis it seems that the flat gluon distribution is slightly preferred over the peaked gluon density.  Including the contribution of resolved photons further improves the agreement with the data, as can be expected, because for a large fraction of the dijet event sample the jet transverse momentum squared is actually larger than the virtuality of the photon ($p_T^2 > Q^2$).

The dijet event sample also provides an independent cross-check of the measurement of the pomeron intercept $\alpha_{\pom}(0)$, as obtained from the Regge-QCD fits to the inclusive cross section.  Figure \ref{fig:jets_xpom_partpom} displays the $\xpom$ distribution of which the slope is directly related to $\alpha_{\pom}(0)$.  Due to the dijet requirement, the kinematic constraints are rather strict and the distribution is seen to rise instead of fall with increasing $\xpom$.  The distribution is, however, still very sensitive to a change in the value of the pomeron intercept, as can be seen from the different Monte Carlo curves corresponding to a range in $\alpha_{\pom}(0)$ between 1.08 and 1.4. Figure \ref{fig:jets_xpom_partpom} also shows the contribution from secondary reggeon exchanges which is small and only visible at large $\xpom$.  Note that the beta distribution extends to much lower values than in the analysis of the inclusive cross section, where $\beta$-values down to 0.04 are reached.

\begin{figure}
\begin{center}
\epsfig{file=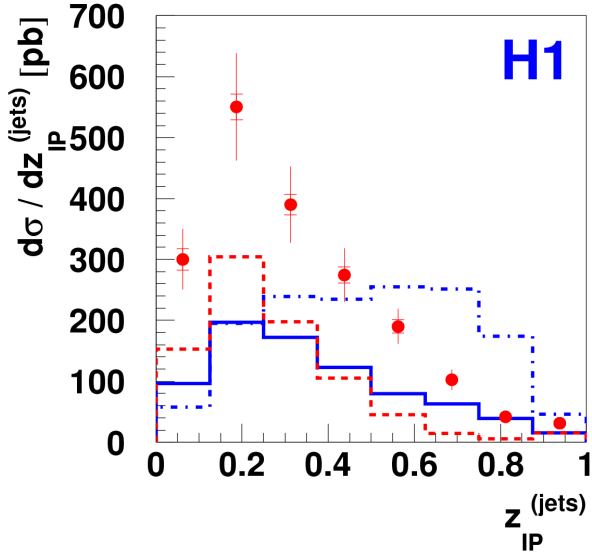,width=0.45\textwidth}
\end{center}
\caption{The differential dijet cross section is displayed as a function of $\zpom$ and compared to different soft colour neutralisations models. The full and dashed-dotted (blue) lines represent the old and new (with general area law) soft colour interaction model; the dashed (red) line represents the semi-classical model.}
\label{fig:jets_zpom_sci}
\end{figure}

Figure \ref{fig:jets_zpom_sci} compares the same experimental data to several soft colour neutralisation models.  The original soft colour interaction model \cite{bib:sci} and the semi-classical model \cite{bib:semiclassical} roughly agree and show the same trend as in the experimental data, although their normalisation is too low by a factor of two.  The new soft colour interaction model (implementing the generalised area law) \cite{bib:sci_gal} has good normalisation, but the shape of the $\zpom$ distribution does not agree with the data.

\begin{figure}
\begin{center}
\epsfig{file=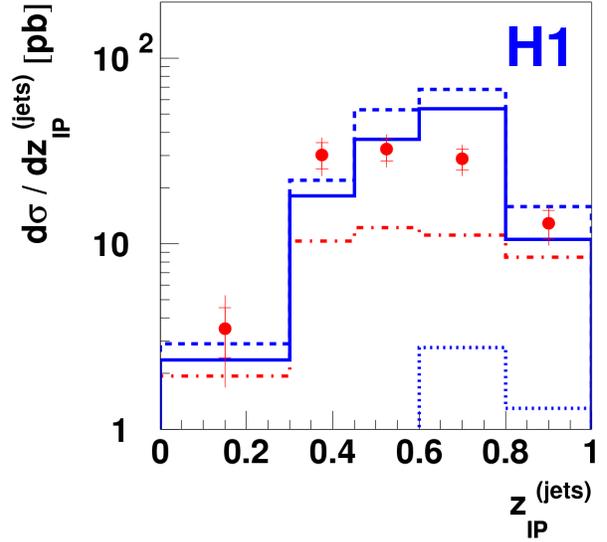,width=0.45\textwidth}
\end{center}
\caption{The differential dijet cross section is displayed as a function of $\zpom$ and compared to the saturation and two-gluon models.  The requirement $\xpom < 0.01$ has been applied in additional to the kinematic cuts of Tab.\@~\ref{tab:jetkine}.  The full, dashed and dotted  (blue) lines are predictions from the two-gluon exchange model for the $q\qbar g$ component with a $p_T^2$ cutoff at $1 {\rm\ GeV}^2$, the $q\qbar g$ component with a $p_T^2$ cutoff at $0.5 {\rm\ GeV}^2$ and the $q\qbar$ component, respectively.  The dashed-dotted (red) line represents the saturation model.}
\label{fig:jets_zpom_2gluon}
\end{figure}

Finally, Fig.\@~\ref{fig:jets_zpom_2gluon} compares the $\zpom$ distribution obtained from data with the saturation model \cite{bib:saturation_model} and two-gluon exchange model \cite{bib:twogluon}.  Here an additional cut on $\xpom$ has been applied  to avoid the valence quark region and to exclude sub-leading exchanges.  The saturation model is somewhat low while the two-gluon exchange model roughly describes the data.  As shown, the $q\qbar$ component of the two-gluon model is nearly negligible and the cross section is by far dominated by $q\qbar g$ partonic states. The two-gluon model requires a cutoff at low transverse momenta of the gluon in $q\qbar g$ states.  Lowering this cutoff to 0.5 significantly increases the cross section and affects mostly the low $p_T$ pomeron remnant.  The model is not able to describe data at $\xpom > 0.01$.

\subsection*{Diffractive open charm production}

The production of charm quarks in diffractive processes provides another test of the underlying dynamics of diffraction models \cite{bib:mcdermott_charm}.  Here, it is the large charm quark mass that makes QCD calculations possible. Moreover, open charm production is free of uncertainties relating to bound states, like in exclusive heavy vector meson production (e.g.\@ $ep \rightarrow e J/\psi p$).  Just as for dijets, charm production is highly sensitive to the role of gluons in the diffractive exchange.

Open charm production is studied through the production of $D^{\ast\pm}$ mesons. Two different decay channels have been used:
\begin{alignat}{2}
D^{\ast +} &\rightarrow D^0 \pi^+, & \qquad D^0 &\rightarrow K^- \pi^+; \\
D^{\ast +} &\rightarrow D^0 \pi^+, & \qquad D^0 &\rightarrow K^- \pi^+ \pi^- \pi^+;
\end{alignat}
with branching ratios of resp.\@ 2.63\% and 5.19\%.  Here and in the following the charge conjugate channels are always implied.

\begin{table}
\begin{center}
\let\PBS=\PreserveBackslash
\setlength{\extrarowheight}{2pt}
\begin{tabular}{>{\PBS\centering}m{39mm}|
                >{\PBS\centering}m{39mm}}
H1 & ZEUS \\
\hline
$2 < Q^2 < 100 {\rm\ GeV}^2$ & $4 < Q^2 < 400 {\rm\ GeV}^2$ \\
$0.05 < y < 0.7$ & $0.02 < y < 0.7$ \\
$p_T(D^\ast) > 2 {\rm\ GeV}$ & $1.5 < p_T(D^\ast) < 8 {\rm\ GeV}$ \\
$|\eta(D^\ast)| < 1.5$ & $|\eta(D^\ast)| < 1.5$ \\
$\xpom < 0.04$ & $\xpom < 0.016$ \\
$M_Y < 1.6 {\rm\ GeV}$ & $\beta < 0.8$ \\
$|t| < 1 {\rm\ GeV}^2$ & \\
\end{tabular}
\end{center}
\caption{Kinematic domains used by the H1 and ZEUS collaborations in the analysis of open charm production in diffractive deep inelastic scattering.} \label{tab:charmkine}
\end{table}

\begin{figure}
\begin{center}
\epsfig{file=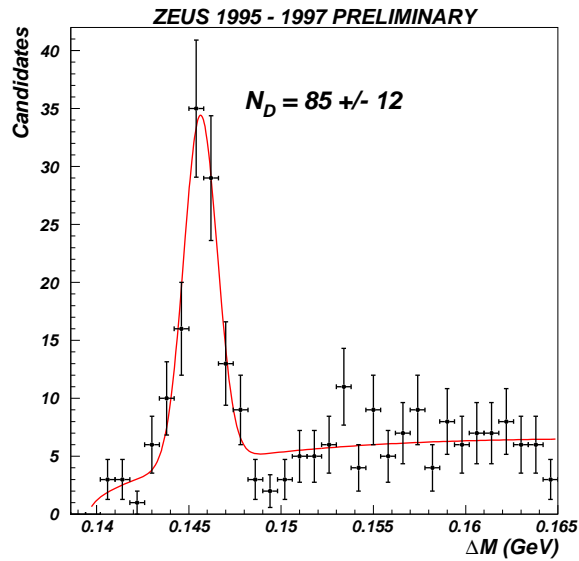,width=0.45\textwidth}
\end{center}
\caption{Distribution of the mass difference $\Delta M = M(K^- \pi^+ \pi^+_{slow}) - M (K^- \pi^+)$.  The curve is a fit of the form $a(\Delta M - M_{\pi^+})^b + {\rm\ Gaussian}$, yielding 85 $D^{\ast\pm}$ candidates.}
\label{fig:charm_mass}
\end{figure}

The ZEUS Collaboration obtained preliminary results using an integrated luminosity of $44 {\rm\ pb}^{-1}$, yielding 85 $D^\ast$ candidates.  The invariant mass spectrum is shown in Fig.\@~\ref{fig:charm_mass}. The cross section is measured in the kinematic domain listed in Tab.\@~\ref{tab:charmkine} and amounts to:
\begin{equation}
\sigma^{ZEUS(prel.)}(ep \rightarrow e D^{\ast\pm} Xp) = 281 \pm 41 ^{+79}_{-73} {\rm\ pb}
\end{equation}

\begin{figure*}
\sidecaption
\epsfig{file=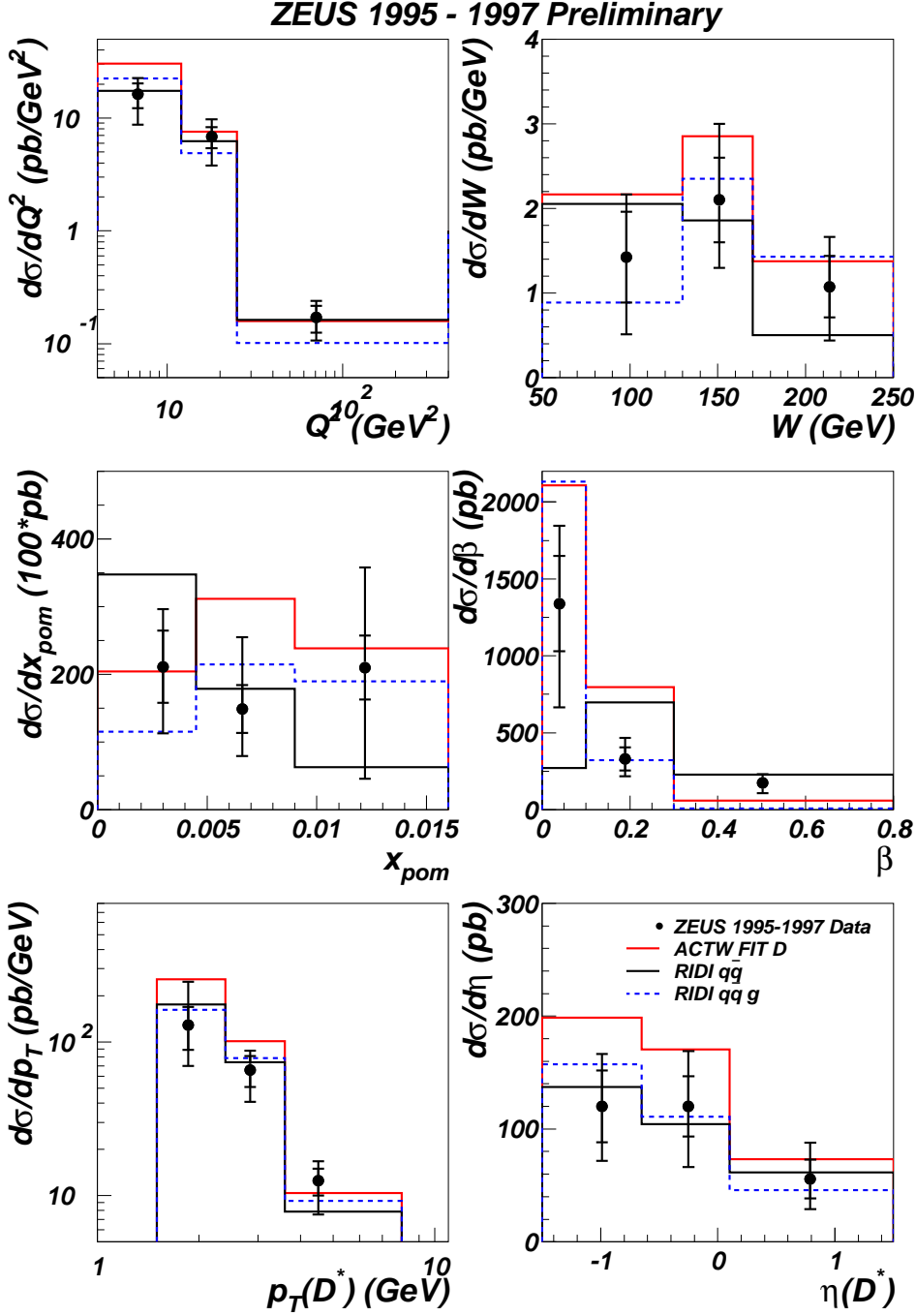,width=0.75\textwidth}
\caption{Differential cross sections for the reaction $ep \rightarrow e D^{\ast\pm} X p$ are plotted as a function of various observables. The $D^{\ast\pm}$ pseudo-rapidity $\eta^{D^{\ast\pm}}$ and transverse momentum $p_T^{D^{\ast\pm}}$ are defined relative to the $\gamma^\ast$ axis in the rest frame of $X$.  The full (black) and dashed (blue) lines are predictions from the two-gluon exchange model and represent the $q \qbar$ and $q \qbar g$ component, respectively.  The dotted (red) curve is the prediction of the partonic pomeron model where the flat gluon distribution (``H1 fit 2'') is used as input.}
\label{fig:charm_xsec}
\end{figure*}

Figure \ref{fig:charm_xsec} shows the differential cross as a function of various observables.  In general, the partonic pomeron and two-gluon models give a reasonable description of the data.  From the $\beta$ spectrum however, it is clear that one needs to combine the $q\qbar$ and $q\qbar g$ components of the two-gluon model in order to describe the full distribution.  The partonic pomeron model seems to yield a $\beta$ distribution that is too steep.

H1 also obtained preliminary results on $D^{\ast\pm}$ production based on an integrated luminosity of $21 {\rm\ pb}^{-1}$ \cite{bib:h1_future_charm}.  Due to differences in detector acceptances, the kinematic domain (listed in Tab.\@~\ref{tab:charmkine}) of the cross section measurement is, however, quite different from the one used by ZEUS, making direct comparison between the two experiments difficult.  The cross section as measured by H1 amounts to:
\begin{equation}
\sigma^{H1(prel.)}(ep \rightarrow e D^{\ast\pm} XY) = 154 \pm 40 \pm 35 {\rm\ pb}
\end{equation}
Although the differential cross sections obtained by H1 and the models agree in shape, here, the normalisation of the overall cross section is very different.  

\begin{table}
\begin{center}
\setlength{\extrarowheight}{2pt}
\begin{tabular}{c|cc}
& H1 & ZEUS \\
\hline
experiment & $154 \pm 40 \pm 35 {\rm\ pb}$ & $281 \pm 41 ^{+79}_{-73} {\rm\ pb}$ \\
RG $\pom$ only & $466 \pm 12 {\rm\ pb}$ & $309 \pm 9 {\rm\ pb}$ \\
RG $\pom+\reg$ & $504 \pm 15 {\rm\ pb}$ & $320 \pm 12 {\rm\ pb}$ \\
\end{tabular}
\end{center}
\caption{Comparison of the measured and predicted cross sections in the kinematic domains as used by H1 and ZEUS.  The prediction are based on the RAPGAP Monte Carlo model with parameters as explained in the text.}
\label{tab:h1_zeus_comp}
\end{table}

Because of the different conclusion reached by H1 and ZEUS, the need for a comparison between experimental results is evident.  The RAPGAP Monte Carlo has been used to make predictions for the different kinematic domains used by the experiments.  The result listed in Tab.\@~\ref{tab:h1_zeus_comp} are obtained using the partonic pomeron model with ``H1 fit 2'', a charm quark mass of $m_c = 1.35 {\rm\ GeV}$, $\Lambda_{QCD} = 0.25$, 5 flavours and the evolution scale set to $\mu^2 = Q^2 + p_T^2 + 4 m_c^2$.  The contribution of sub-leading reggeon exchange is added separately.  The RAPGAP predictions are in agreement with the ZEUS results, but are a factor three times larger than the H1 result.  The question remains whether this can explained by the inability of the model to bridge the gap between kinematic domains.  It is clear that new and more precise experimental data are necessary to resolve this issue.

\section{Summary}
\label{sec:summary}

Recent preliminary results, obtained by the H1 and ZEUS collaborations, on high $E_T$ dijet production and open charm production in diffractive deep inelastic scattering have been presented and compared to several theoretical models.

The partonic pomeron model, with a flat gluon density function (``H1 fit 2''), produces a very good description of all experimental data.  The dijet analysis in particular thus confirms the picture of a hard pomeron ($\alpha_{\pom}(0) = 1.2$) dominated by gluons carrying large fractional momenta.

The soft colour interaction, semi-classical and saturation models are either wrong in normalisation or in shape.  As some of these models might be improved by including next-to-leading order contributions, it is presently not yet possible to make any final judgements on their validity.

Full perturbative QCD calculations, based on the exchange of two gluons in a colour-singlet configuration, are able to describe the differential dijet cross sections, but are limited to a kinematic domain where $\xpom < 0.01$.

H1 and ZEUS results on open charm production in diffractive deep inelastic scattering suffer from low statistics and are, at the moment, difficult to reconcile.  Until new and more precise data become available, it is not possible to draw any firm conclusions based on these results.

\renewcommand{\ackname}{Acknowledgements\runinend}
\begin{acknowledgement}
I would like to thank the organizers of CRIMEA-2000 for their hospitality and the invitation to this interesting summer school-seminar.  I appreciate the hard work of all members of the H1 and ZEUS collaborations who contributed to these results by collecting and analysing the experimental data.
\end{acknowledgement}

\bibliographystyle{unsrt}
\bibliography{proc}

\begin{thebibliography}{10}

\bibitem{bib:collins_regge_theory}
{P.\@ D.\@ B.\@ Collins,}~{\it Introduction to Regge Theory and High Energy
  Physics,} {Cambridge University Press (1977)}.

\bibitem{bib:dl}
{A.\@ Donnachie and P.\@ V.\@ Landshoff,} {\PL{B296}{1992}{227--232}}.

\bibitem{bib:forshaw_qcdpomeron}
{J.\@ R.\@ Forshaw and D.\@ A.\@ Ross,}~{\it Quantum Chromodynamics and the
  Pomeron,} {Cambridge University Press (1997)}.

\bibitem{bib:hera_lrg_obs}
{\ZC, M.\@ Derrick \ea,}~{\PL{B315}{1993}{481--493},}
  {\ib{B346}{1995}{399--414};} \\ {\HC, T.\@ Ahmed
  \ea,}~{\NP{B429}{1994}{477--502},} {\ib{B435}{1995}{3--22}}.

\bibitem{bib:ingelman_schlein}
{G.\@ Ingelman and P.\@ Schlein,} {\PL{B152}{1985}{256}}.

\bibitem{bib:hera_incl_ddis}
{\HC, T.\@ Ahmed \ea,}~{\PL{B348}{1995}{681--696};} \\ {\ZC, M.\@ Derrick \ea,}
  {\ZP{C68}{1995}{569--584},} {\ib{C70}{1996}{391--412},}~{J.\@ Breitweg \ea,}
  {\EPJ{C1}{1998}{81--96}}.

\bibitem{bib:h1_reggeqcd_fits}
{\HC, C.\@ Adloff \ea,}~{\ZP{C76}{1997}{613--629}}.

\bibitem{bib:zeus_diffdis_ratio}
{\ZC, J.\@ Breitweg \ea,}~{\EPJ{C6}{1999}{43--66}}.

\bibitem{bib:hera_diff_hfs}
{\ZC, M.\@ Derrick \ea,}~{\PL{B338}{1994}{483--496},}
  {\ZP{C67}{1995}{227--238},} {J.\@ Breitweg \ea,} {\PL{B421}{1998}{368--384};}
  \\ {\HC, S.\@ Aid \ea,} {\ZP{C70}{1996}{609--620},} {C.\@ Adloff \ea,}
  {\PL{B428}{1998}{206--220},}~{\EPJ{C1}{1998}{495--507},}
  {\ib{C5}{1998}{439--452}}.

\bibitem{bib:hera_diff_excl_hfs}
{\ZC, M.\@ Derrick \ea,}~{\PL{B332}{1994}{228--243};} \\~{\HC, C.\@ Adloff
  \ea,} {\EPJ{C6}{1999}{421--436}}.

\bibitem{bib:gribov_pomeranchuk}
{V.\@ N.\@ Gribov and I.\@ Ya.\@ Pomeranchuk,} {\JETP{15}{1962}{788L}}.

\bibitem{bib:collins_diff_qcdfact}
{J.\@ C.\@ Collins,}~{\PR{D57}{1998}{3051--3056},} {\ib{D61}{2000}{019902}
  (erratum)}.

\bibitem{bib:diff_pdf}
{L.\@ Trentadue and G.\@ Veneziano,} {\PL{B323}{1994}{201--211};} \\~{A.\@
  Berera and D.\@ E.\@ Soper,} {\PR{D53}{1996}{6162--6179}}.

\bibitem{bib:dglap}
{V.\@ N.\@ Gribov and L.\@ N.\@ Lipatov,} {\SJNP{15}{1972}{438--450 and
  675--684};} \\~{Yu.\@ L.\@ Dokshitzer,} {\JETP{46}{1977}{641--653};} \\ {G.\@
  Altarelli and G.\@ Parisi,}~{\NP{B126}{1977}{298}}.

\bibitem{bib:sci}
{A.\@ Edin, G.\@ Ingelman and J.\@ Rathsman,}~{\PL{B366}{1996}{371--378},}
  {\ZP{C75}{1997}{57--70}}.

\bibitem{bib:sci_gal}
{J.\@ Rathsman,} {\PL{B452}{1999}{364--371}}.

\bibitem{bib:bartels_photonfluct}
{J.\@ Bartels, J.\@ Ellis, H.\@ Kowalski and M.\@ W\"usthoff,}
  {\EPJ{C7}{1999}{443--458}}.

\bibitem{bib:saturation_model}
{K.\@ Golec-Biernat, M.\@ W\"usthoff,} {\PR{D59}{1999}{014017}}.

\bibitem{bib:semiclassical}
{W.\@ Buchm\"uller, M.\@ F.\@ McDermott and A.\@ Hebecker,}
  {\NP{B487}{1997}{283--310};} \\~{W.\@ Buchm\"uller, T.\@ Gehrmann and A.\@
  Hebecker,} {\NP{B537}{1999}{477--500};}.

\bibitem{bib:twogluon}
{J.\@ Bartels, H.\@ Lotter and M.\@ W\"usthoff,} {\PL{B379}{1996}{239--248};}
  \\ {J.\@ Bartels, C.\@ Ewerz, H.\@ Lotter and M.\@ W\"usthoff,}
  {\PL{B386}{1996}{389--396};} \\~{J.\@ Bartels, H.\@ Jung, M.\@ W\"usthoff,}
  {\EPJ{C11}{1999}{111--125}}.

\bibitem{bib:rapgap}
{H.\@ Jung,} {\CPC{86}{1995}{147--161}}.

\bibitem{bib:lepto}
{A.\@ Edin, G.\@ Ingelman and J.\@ Rathsman,} {\CPC{101}{1997}{108--134}}.

\bibitem{bib:mcdermott_dijets}
{W.\@ Buchm\"uller, M.\@ F.\@ McDermott and A.\@ Hebecker,}
  {\PL{B410}{1997}{304--310}}.

\bibitem{bib:h1_future_dijets}
{F.-P.\@ Schilling,} {Ph.\@ D.\@ thesis,}~{University of Heidelberg,} {to be
  published}.

\bibitem{bib:mcdermott_charm}
{W.\@ Buchm\"uller, M.\@ F.\@ McDermott and A.\@ Hebecker,}
  {\PL{B404}{1997}{353--361}}.

\bibitem{bib:h1_future_charm}
{P.\@ Thompson,}~{Ph.\@ D.\@ thesis,} {University of Birmingham}.

\end{thebibliography}

\end{document}